\documentclass[12pt]{article}
\usepackage{epsfig}
\usepackage{axodraw}
\usepackage{epsfig}
\usepackage{psfrag}
\usepackage{cite}
\usepackage{graphicx}
\usepackage{rotate}
\usepackage{latexsym}
\usepackage{amssymb}
\newcommand\pubnumber{
NYU-TH/03-07-27\\
Freiburg-THEP 03/09
}

\newcommand\hepnumber{hep-ph/0307200}

\def\csuma{Fakult\"at f\"ur Physik,  Universit\"at Freiburg,\\ D-79104 Freiburg, Germany\\ and\\ 
Institut f\"ur Theoretishe Theilchenphysik, Universit\"at
Karlsrhue,\\
D-76128 Karlsruhe, Germany}
\def\csumb{Department of Physics, New York University, \\ 4 Washington Place, New York, NY 10003, USA}

\def\Title#1{\begin{center} {\Large\bf #1 } \end{center}}
\def\Author#1{\begin{center}{ \sc #1} \end{center}}
\def\Address#1{\begin{center}{ \it #1} \end{center}}

\newcommand\pubblock{\rightline{\begin{tabular}{l} \pubnumber\\
\hepnumber \end{tabular}}}
\newenvironment{Abstract}{\begin{quotation}  }{\end{quotation}}

\makeatletter
\def\section{\@startsection{section}{0}{\z@}{5.5ex plus .5ex minus
 1.5ex}{2.3ex plus .2ex}{\large\bf}}
\def\subsection{\@startsection{subsection}{1}{\z@}{3.5ex plus .5ex minus
 1.5ex}{1.3ex plus .2ex}{\normalsize\bf}}
\def\subsubsection{\@startsection{subsubsection}{2}{\z@}{-3.5ex plus
-1ex minus  -.2ex}{2.3ex plus .2ex}{\normalsize\sl}}

\renewcommand{\@makecaption}[2]{%
   \vskip 10pt
   \setbox\@tempboxa\hbox{\small #1: #2}
   \ifdim \wd\@tempboxa >\hsize     
       \small #1: #2\par          
     \else                        
       \hbox to\hsize{\hfil\box\@tempboxa\hfil}
   \fi}
 \def\citenum#1{{\def\@cite##1##2{##1}\cite{#1}}}
\def\citea#1{\@cite{#1}{}}
\newcount\@tempcntc
\def\@citex[#1]#2{\if@filesw\immediate\write\@auxout{\string\citation{#2}}\fi
  \@tempcnta\z@\@tempcntb\m@ne\def\@citea{}\@cite{\@for\@citeb:=#2\do
    {\@ifundefined
       {b@\@citeb}{\@citeo\@tempcntb\m@ne\@citea\def\@citea{,}{\bf }\@warning
       {Citation `\@citeb' on page \thepage \space undefined}}%
    {\setbox\z@\hbox{\global\@tempcntc0\csname b@\@citeb\endcsname\relax}%
     \ifnum\@tempcntc=\z@ \@citeo\@tempcntb\m@ne
       \@citea\def\@citea{,}\hbox{\csname b@\@citeb\endcsname}%
     \else
      \advance\@tempcntb\@ne
      \ifnum\@tempcntb=\@tempcntc
      \else\advance\@tempcntb\m@ne\@citeo
      \@tempcnta\@tempcntc\@tempcntb\@tempcntc\fi\fi}}\@citeo}{#1}}
\def\@citeo{\ifnum\@tempcnta>\@tempcntb\else\@citea\def\@citea{,}%
  \ifnum\@tempcnta=\@tempcntb\the\@tempcnta\else
  {\advance\@tempcnta\@ne\ifnum\@tempcnta=\@tempcntb \else\def\@citea{--}\fi
    \advance\@tempcnta\m@ne\the\@tempcnta\@citea\the\@tempcntb}\fi\fi}
\makeatother
\newcommand{\bmath}{\begin{displaymath}}
\newcommand{\emath}{\end{displaymath}}
\newcommand{\be}{\begin{equation}}
\newcommand{\ee}{\end{equation}}
\newcommand{\bea}{\begin{eqnarray}}
\newcommand{\eea}{\end{eqnarray}}

\newcommand{\kee}{\hat{\kappa}^{(e;\, e)}}
\newcommand{\seff}{\sin^2 \theta_{eff}^{lept}}
\newcommand{\seq}{\sin^2 \hat{\theta}_W (q^2)}
\newcommand{\sms}{\hat{s}^2}
\newcommand{\cms}{\hat{c}^2}
\newcommand{\kpt}{\hat{\kappa}^{PT}(q^2, \mu)}
\newcommand{\kptm}{\hat{\kappa}^{PT}(q^2, m_Z)}
\newcommand{\see}{\sin^2 \theta_{eff}^{(e;\, e)}(q^2)}
\newcommand{\seezero}{\sin^2 \theta_{eff}^{(e;\, e)}(0)}
\begin{document}
\begin{titlepage}
\pubblock
\vfill
\def\thefootnote{\fnsymbol{footnote}}
\boldmath \Title{The electroweak form factor $\hat{\kappa}(q^2)$ 
\\
and
the running of $\sin^2\hat{\theta}_W$} \unboldmath
\vfill
\Author{A.~Ferroglia$^1$\footnote{\tt{andrea.ferroglia@physik.uni-freiburg.de}},
G.~Ossola$^2$\footnote{\tt{giovanni.ossola@physics.nyu.edu}}, and
A.~Sirlin$^2$\footnote{\tt{alberto.sirlin@nyu.edu}}}
\Address{$^1$\csuma}
\Address{$^2$\csumb}
\vfill
\begin{Abstract} 
Gauge independent form factors $\rho^{(e ;\, e)}$ and
$\hat{\kappa}^{(e ;\, e)}(q^2)$ for M\o ller scattering at \mbox{$s \ll
m_{W}^2$} are derived. It is pointed out that $\kee$ is very
different from its counterparts in other processes. The relation
between the effective parameter $\kee(q^2,\mu)\,\sin^2
\hat{\theta}_{W}(\mu)$ and $\seff$ is derived in a  scale-independent 
manner. A gauge and process-independent running
parameter $\sin^2 \hat{\theta}_{W}(q^2)$, based on the pinch-technique 
self-energy $a_{\gamma Z} (q^2)$, is discussed for all
$q^2$ values. At $q^2=0$ it absorbs very accurately the
Czarnecki-Marciano calculation of the M\o ller scattering
asymmetry at low $s$ values, and at $q^2 = m_{Z}^2$ it is rather
close to $\seff$. The $q^2$ dependence of $\seq$ is displayed in
the space and time-like domains.
\end{Abstract}


\end{titlepage}
\def\thefootnote{\arabic{footnote}}
\setcounter{footnote}{0}

\section{Introduction}
The form factors $\rho$ and $\kappa(q^2)$ that incorporate the effect
of the electroweak corrections in the neutral current amplitudes
have played an important role in precision studies of the Standard
Model (SM) \cite{n1}. 
In particular, their effect has been discussed in detail
in $\nu$-hadron and $\nu$-lepton scattering at momentum
transfers $|q^2| \ll m_{W}^2$ \cite{m1,m2,m3}, as well as in $e^+ + e^-
\rightarrow f + \overline{f}$ near the $Z^0$ peak \cite{m4}. We
recall that $\kappa(q^2)$ accompanies the electroweak mixing parameter
$\sin^2 \theta_{W}$ in the $Z^0-f\overline{f}$ coupling, while
$\rho$ multiplies the full neutral current amplitude. A modified
version of $\kappa(m_Z^2)$, denoted as $\hat{k}(m_Z^2)$, has also been
important in establishing the connection between $\seff$, employed
by the Electroweak Working Group (EWWG) to analyze physics at
the $Z^0$ peak, and the $\overline{\mbox{MS}}$ parameter $\sin^2
\hat{\theta}_W (m_Z)$ \cite{m5}. The amplitude $\kappa(q^2)$ and the
parameter $\sin^2 \theta_W$ are frequently  used in two
renormalization schemes: the on-shell framework where $\sin^2
\theta_W \equiv 1- m_W^2/m_Z^2$  \cite{m6}, and the $\overline{\mbox{MS}}$
approach, where one employs $\sin^2 \hat{\theta}_W (\mu)$ and the
form factor is denoted as $\hat{\kappa}$. One has, by definition, the
relation \cite{m2}
\be \kappa(q^2) \sin^2 \theta_W = \hat{\kappa}(q^2, \mu) \sin^2 \hat{\theta}
(\mu)\,, \ee
where $\mu$ is the 't~Hooft scale. In the on-shell scheme,
$\kappa(q^2)$ and $\sin^2 \theta_W$ are $\mu$-independent and can be
separately regarded as physical observables, while in the
$\overline{\mbox{MS}}$ framework it is the combination
$\hat{\kappa}(q^2, \mu) \sin^2 \hat{\theta} (\mu)$ that plays that
function. The traditional construction of $\rho$ and
$\hat{\kappa}(q^2,\mu)$ at $|q^2| \ll m_W^2$ leads to gauge invariant
expressions, which are, however, process dependent.

Recently, Czarnecki and Marciano \cite{m7,m8,m9} emphasized that $\hat{\kappa}$ is
particularly important in polarized M\o ller scattering at low $s$
values ($s = (p_1 + p_2)^2$ where $p_1$ and $p_2$ are the four
momenta of the initial electrons). In fact, at the tree-level the
asymmetries measured in that process are proportional to $1 - 4
\sin^2 \hat{\theta}_W$, a very small number. Since in the presence
of electroweak corrections this factor is replaced at $q^2 =0$ by
$1 - 4 \kee(0, m_Z) \sin^2 \hat{\theta}_W (m_Z)$, and $\kee(0, m_Z) \approx
1.03$, their effect induces a sharp reduction in the predicted
asymmetries. This observation is of particular interest at present in view of the 
proposed E158 fixed target experiment at SLAC \cite{e158}. 

In Section 2 we discuss the construction of gauge independent form
factors $\rho^{(e ;\, e)}$ and $\kee(q^2,\mu)$ relevant to M\o
ller scattering at $s$ (and therefore $q^2$) $\ll m_W^2$, and
evaluate them in the framework of the general $R_\xi$ gauge. Their
comparison with the traditionally defined amplitudes
for other processes, such as $\nu$-lepton
scattering at $|q^2| \ll m_W^2$, is discussed.
The relation between the effective angle defined by $\kee(q^2,\mu)\,
\sin^2 \hat{\theta}_W (\mu)$ and $\seff$ is analyzed.

In Section 3 we discuss a gauge and process independent running
parameter $\sin^2 \hat{\theta}_W(q^2)$, defined
on the basis of the Pinch Technique (PT) $\gamma - Z$ self-energy,
and show that it approximates very accurately the electroweak
corrections for M\o ller scattering at low $s$ values, while it is rather close to
$\seff$ at $q^2 = m_Z^2$. The $q^2$ dependence of $\sin^2 \hat{\theta}_W (q^2)$ 
is then illustrated over a large range of values in the space-like and time-like domains.

\boldmath
\section{Gauge-Independent Form Factors $\rho^{(e ;\, e)}$ and
$\kee(q^2, \mu)$} \unboldmath

The traditional approach to obtain the form factors $\rho$ and
$\hat{\kappa}$ at $|q^2| \ll m_W^2$ in the case of $\nu - l$ and $\nu$-
hadron scattering \cite{m2,m3}, has been to consider the electroweak
corrections in the limit in which external fermion masses and
momentum transfers are neglected relative to $m_W$. The electroweak
corrections can then be written as expressions bilinear in the
matrix elements of $J_\mu^Z$ and $J_\mu^\gamma$, the fermionic
currents coupled to $Z^0$ and $\gamma$, respectively.
Contributions bilinear in $J_\mu^Z$ are absorbed in $\rho$
while those proportional to $J_\mu^Z \, J^\mu_\gamma$ define the
form factor $\hat{\kappa}(q^2)$. Photonic  corrections to the external
legs are treated separately. Writing the fermionic current of $Z^0$
in the form
\be J_{\mu}^Z =
\overline{\psi}\, \frac{C_3\, \gamma_\mu\, a_{-}\, }{2}\, \psi - \sin^2
\hat{\theta}_W\,J_\mu^\gamma \, , \ee
where $a_- \equiv (1 - \gamma_5)/2$,  $J_{\mu}^\gamma$ is the
electromagnetic current and $C_3$ ($\pm 1$) is twice the third
component of weak isospin, this procedure leads to the
replacement of the tree-level neutral current amplitude $M_{Z
Z}^0$ by
\be M^0_{Z Z} \rightarrow \frac{i m_Z^2}{q^2 - m_Z^2}\,\frac{8
G_F}{\sqrt{2}}\,\rho^{(i ;\, i')}\,<f| J_\mu^Z |i>\,<f'| J_Z^\mu
|i'> \, , \label{eq3}\ee
\be J_\mu^Z \rightarrow
\overline{\psi}\, \frac{C_3\, \gamma_\mu\, a_{-}\, }{2}\, \psi -
\hat{\kappa}^{(i ;\, i')}(q^2, \mu)\,\sin^2
\hat{\theta}_W (\mu)\,J_\mu^\gamma \, , \label{eq3bis}\ee
where $G_F = 1.16637(1) \times 10^{-5}/ \mbox{GeV}^2$ is the Fermi
constant determined from $\mu$ decay, and the superscripts $(i
;\, i')$ refer to the initial fermions in the process under
consideration. Thus, as mentioned in Section 1, $\rho$ and
$\hat{\kappa}$ are process dependent. In some cases, like $\nu-l$
scattering, it is possible to absorb the complete electroweak
corrections in these form factors. In other cases, such as
$\nu$-hadron scattering, this is not possible since the
electroweak corrections induce hadronic currents with an isospin
structure not present at the tree level. The latter are then
treated as additional contributions to those contained in
Eqs.~(\ref{eq3},\ref{eq3bis}).

The approach described above has two important virtues: i) the
dominant electroweak corrections are incorporated as compact and
rather simple modifications  of the tree-level neutral current
amplitude and ii) the form factors $\rho^{(i ;\, i')}$ and
$\hat{\kappa}^{(i ;\, i')}(q^2, \mu)$ are gauge independent.

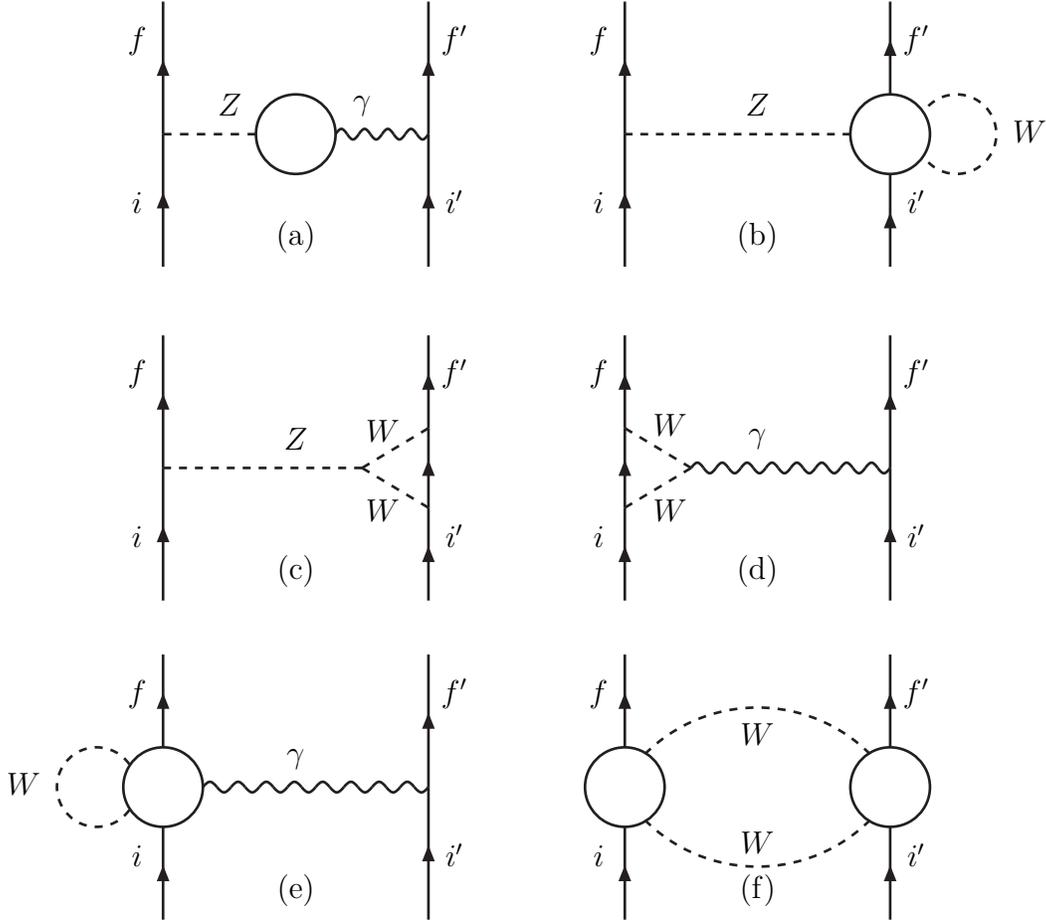
\begin{figure}[h]
\vspace*{-8mm}
\[
  \vcenter{
\hbox{
  \begin{picture}(0,0)(0,0)
  \SetWidth{1.}
   \ArrowLine(-50,-50)(-50,0)
\ArrowLine(-50,0)(-50,50)
    \DashLine(-50,0)(-15,0){3}
    \Photon(15,0)(50,0){2}{4}
    \CArc(0,0)(15,-180,180)   
	\ArrowLine(50,-50)(50,0)
\ArrowLine(50,0)(50,50)
    \Text(-60,-30)[cb]{$i$}
    \Text(60,-30)[cb]{$i'$}
    \Text(-60,30)[cb]{$f$}
    \Text(60,30)[cb]{$f'$} 
    \Text(0,-45)[cb]{(a)}
    \Text(-25,7)[cb]{$Z$}
    \Text(25,7)[cb]{$\gamma$} 
\end{picture}}
}
\hspace{6cm}
  \vcenter{
\hbox{
  \begin{picture}(0,0)(0,0)
  \SetWidth{1.}
   \ArrowLine(-50,-50)(-50,0)
\ArrowLine(-50,0)(-50,50)
    \DashLine(-50,0)(35,0){3}
    \DashCArc(75,0)(15,-135,135){3}
    \CArc(50,0)(15,-180,180)   
	\ArrowLine(50,-50)(50,-15)
\ArrowLine(50,15)(50,50)
    \Text(-60,-30)[cb]{$i$}
    \Text(60,-30)[cb]{$i'$}
    \Text(-60,30)[cb]{$f$}
    \Text(60,30)[cb]{$f'$} 
    \Text(0,-45)[cb]{(b)}
    \Text(0,7)[cb]{$Z$} 
    \Text(103,-3)[cb]{$W$}
\end{picture}}
}
\]
\vspace{3cm}
\[
  \vcenter{\hbox{
  \begin{picture}(0,0)(0,0)
  \SetWidth{1.}
   \ArrowLine(-50,-50)(-50,0)
\ArrowLine(-50,0)(-50,50)
    \DashLine(-50,0)(25,0){3}
    \DashLine(25,0)(50,-15){3}
     \DashLine(25,0)(50,15){3}  
	\ArrowLine(50,-50)(50,-15)
\ArrowLine(50,-15)(50,15)
\ArrowLine(50,15)(50,50)
    \Text(-60,-30)[cb]{$i$}
    \Text(60,-30)[cb]{$i'$}
    \Text(-60,30)[cb]{$f$}
    \Text(60,30)[cb]{$f'$} 
    \Text(0,-45)[cb]{(c)}
    \Text(0,7)[cb]{$Z$} 
    \Text(33,10)[cb]{$W$}
    \Text(33,-20)[cb]{$W$}
    
\end{picture}}}
\hspace{6cm}
  \vcenter{\hbox{
  \begin{picture}(0,0)(0,0)
  \SetWidth{1.}
   \ArrowLine(50,-50)(50,0)
\ArrowLine(50,0)(50,50)
    \Photon(-25,0)(50,0){2}{8}
    \DashLine(-25,0)(-50,-15){3}
     \DashLine(-25,0)(-50,15){3}  
	\ArrowLine(-50,-50)(-50,-15)
\ArrowLine(-50,-15)(-50,15)
\ArrowLine(-50,15)(-50,50)
    \Text(-60,-30)[cb]{$i$}
    \Text(60,-30)[cb]{$i'$}
    \Text(-60,30)[cb]{$f$}
    \Text(60,30)[cb]{$f'$} 
    \Text(0,-45)[cb]{(d)}
    \Text(0,7)[cb]{$\gamma$} 
    \Text(-33,12)[cb]{$W$}
    \Text(-33,-20)[cb]{$W$}
    
\end{picture}}}
\]
\vspace{3cm}
\[
  \vcenter{\hbox{
  \begin{picture}(0,0)(0,0)
  \SetWidth{1.}
   \ArrowLine(50,-50)(50,0)
\ArrowLine(50,0)(50,50)
    \Photon(50,0)(-35,0){2}{8}
    \DashCArc(-75,0)(15,35,-35){3}
    \CArc(-50,0)(15,-180,180)   
	\ArrowLine(-50,-50)(-50,-15)
\ArrowLine(-50,15)(-50,50)
    \Text(-60,-30)[cb]{$i$}
    \Text(60,-30)[cb]{$i'$}
    \Text(-60,30)[cb]{$f$}
    \Text(60,30)[cb]{$f'$} 
    \Text(0,-45)[cb]{(e)}
    \Text(0,7)[cb]{$\gamma$} 
    \Text(-103,-3)[cb]{$W$}
\end{picture}}}
\hspace{6cm}
  \vcenter{\hbox{
  \begin{picture}(0,0)(0,0)
  \SetWidth{1.}
  
    \DashCArc(0,-30)(60,45,135){3}
    \DashCArc(0,30)(60,-135,-45){3}
    \CArc(-50,0)(15,-180,180) 
\CArc(50,0)(15,-180,180)  
	\ArrowLine(-50,-50)(-50,-15)
\ArrowLine(-50,15)(-50,50)
	\ArrowLine(50,-50)(50,-15)
\ArrowLine(50,15)(50,50)
    \Text(-60,-30)[cb]{$i$}
    \Text(60,-30)[cb]{$i'$}
    \Text(-60,30)[cb]{$f$}
    \Text(60,30)[cb]{$f'$} 
    \Text(0,-45)[cb]{(f)}
    \Text(0,-25)[cb]{$W$} 
    \Text(0,16)[cb]{$W$}
\end{picture}}}
\]
\vspace{2cm}
\caption[]{ Feynman diagrams that give gauge-dependent contributions to the 
$\hat{\kappa}$ form factor in \mbox{$<f'| J_\mu^Z |i'>$}. 
The circle in Fig.~(1a) represents the contribution of $A_{Z \gamma}$, 
including its fermionic and bosonic components. The solid line circles in Figs.~(1b,1e)
indicate a sum of diagrams in which the ends of the $W$ propagator are attached
in all possible ways to the fermion lines. In particular, they include the $W$ contribution
to the wave function renormalization of the external lines. Diagram (1f) represents
the $W$-$W$ box diagrams, whether crossed or uncrossed.}
\label{fig1}
\end{figure}

In order to implement this construction in the case of  M\o
ller scattering, we consider the diagrams of Fig.~\ref{fig1},
which contain all the gauge-dependent contributions to the
$\hat{\kappa}$ form factor associated with the $<f'| J_\mu^Z |i'>$
matrix element. The mirror image of those diagrams contributes to the
$\hat{\kappa}$ factor present in the $<f| J_\mu^Z |i>$ amplitude.

Using the results of Ref.~\cite{m10}, the diagrams of
Fig.~\ref{fig1} can be calculated in the general $R_\xi$ gauge.
The gauge dependencies of the graphs indeed cancel and, performing
the $\overline{\mbox{MS}}$ subtraction, we find that their
contribution to $\hat{\kappa}$ in the case of the M\o ller scattering at $s$
(and therefore $|q^2|$) $\ll m_W^2$ is given by
\bea \kee(q^2, \mu) &=& 1 + \frac{\alpha}{2 \pi
\hat{s}^2}\,\ln{\left(\frac{m_Z}{\mu}\right)}\left[-\frac{1}{3}\,\sum_i\,(C_i\,Q_i
-4\,\sms\,Q_i^2) + 7\,\cms +\frac{1}{6}\right] \nonumber \\ &+&
\frac{\alpha}{2 \pi \hat{s}^2}\,\left[- \sum_i (C_i\,Q_i
-4\,\sms\,Q_i^2)\,I_i(q^2)+ \left(\frac{7}{2}\,\cms +
\frac{1}{12}\right)\,\ln{c^2} - \frac{23}{18} +\frac{\sms}{3}
\right]\, ,\label{eq5}
 \eea
where
\be I_i (q^2) = \int_0^1 dx\, x\,(1-x)\,\ln{\frac{m_i^2 -
q^2\,x\,(1-x)}{m_Z^2}}\, , \ee
the $i$ summation is over the fundamental fermions and includes a
color factor $3$ for quarks, $C_i$ ($\pm 1$) is twice the
third component of weak isospin for fermion $i$, $Q_i$ is its
electric charge in units of the proton charge $e$, $m_i$ its mass, $c^2 \equiv
m_W^2 / m_Z^2$, and $\sms \equiv 1 -\cms$ is an abbreviation for
the $\overline{\mbox{MS}}$ parameter $\sin^2 \hat{\theta}_W
(m_Z)$.

The corresponding gauge independent form factor $\rho^{(e ;\, e)}$
contains contributions from several diagrams discussed in
Ref.~\cite{m1}, and in the M\o ller scattering case becomes
\be \rho^{(e ;\, e)} = 1 +\frac{\hat{\alpha}(m_Z)}{4 \pi
\sms}\,\left\{\frac{3}{4}\,\frac{\ln{c^2}}{\sms}  - \frac{3}{4} +
\frac{3}{4}\,\frac{m_t^2\,(1 + \delta_{QCD})}{m_W^2} +
\frac{3}{4}\,\xi\,\left(\frac{\ln{(c^2/\xi)}}{c^2 - \xi} +
\frac{1}{c^2}\,\frac{\ln{\xi}}{1 -\xi}\right) \right\} \,
,\label{eq6}\ee
where $\xi \equiv m_H^2 /m_Z^2$, $\hat{\alpha}^{-1}(m_Z) = 1/127.9
\pm 0.1$,  and $\delta_{QCD} \approx - 0.12$ is a  QCD correction
\cite{m11}.

\begin{figure}[h]
\vspace{3cm}
\[
  \vcenter{\hbox{
  \begin{picture}(0,0)(0,0)
  \SetWidth{1.}
   \ArrowLine(50,-50)(50,0)
\ArrowLine(50,0)(50,50)
    \Photon(50,0)(-35,0){2}{8}
    \DashCArc(-75,0)(15,35,-35){3}
    \CArc(-50,0)(15,-180,180)   
	\ArrowLine(-50,-50)(-50,-15)
\ArrowLine(-50,15)(-50,50)
    \Text(-60,-30)[cb]{$i$}
    \Text(60,-30)[cb]{$i'$}
    \Text(-60,30)[cb]{$f$}
    \Text(60,30)[cb]{$f'$} 
    \Text(0,7)[cb]{$\gamma$} 
    \Text(-103,-3)[cb]{$Z$}
\end{picture}}}
\]
\vspace{2cm}
\caption[]{Gauge-independent vertex diagram. The meaning of the 
solid line circle is the same as in Fig.~\ref{fig1}.}
\label{fig2}
\end{figure}
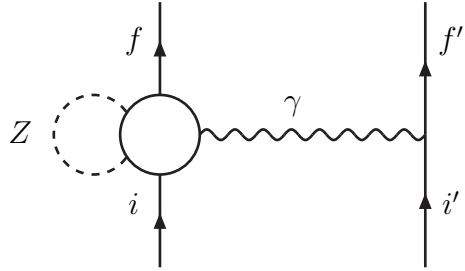

Aside from the electroweak corrections in Eqs.~(\ref{eq5},
\ref{eq6}), there are several additional contributions involving
$Z-Z$ and $\gamma-Z$ boxes, the vertex diagram of Fig.~\ref{fig2},
and QED corrections. These additional contributions are gauge
independent and have been evaluated separately in Ref.~\cite{m7}.

At $q^2 =0$, Eq.~(\ref{eq5}) becomes
\bea \kee(0, \mu) &=& 1 + \frac{\alpha}{2 \pi
\hat{s}^2}\,\ln{\left(\frac{m_Z}{\mu}\right)}\left[- \frac{1}{3}\,\sum_i\,(C_i\,Q_i
-4\,\sms\,Q_i^2) + 7\,\cms +\frac{1}{6}\right] \nonumber \\ &+&
\frac{\alpha}{2 \pi \hat{s}^2}\,\left[\frac{1}{3}\,\sum_i
(C_i\,Q_i -4\,\sms\,Q_i^2)\,\ln{\left(\frac{m_Z}{m_i} \right)}+
\left(\frac{7}{2}\,\cms + \frac{1}{12}\right)\,\ln{c^2} - \frac{23}{18}
+\frac{\sms}{3} \right]\, .\label{eq7}
 \eea
In order to evaluate $\kee(0, \mu)\,\sin^2 \hat{\theta}_W(\mu)$ it
is convenient to set $\mu =m_Z$. The parameter $\sin^2
\hat{\theta}(m_Z)$ can be obtained from $\seff$ by using the
analysis of Ref.~\cite{m5}. Since in Section~3 we will consider a
running parameter at large $q^2$ values, for the purpose of this
paper we choose not to decouple the top quark in the explicit
summations in Eqs.~(\ref{eq5},\ref{eq7}), or in the definition of
$\sin^2 \hat{\theta}(m_Z)$. In that case, for $m_t = 174.3\ \mbox{GeV}$, we have  
$\sin^2 \hat{\theta}(m_Z)
= \seff/ 1.00044$. 
For $\seff$ we will use the
central value of the current world average, 0.23148 \cite{m12}, which leads to 
$\sin^2 \hat{\theta}(m_Z) = 0.23138$. In the Appendix we report the results obtained 
if, instead, one employs as input the central value 
derived from the leptonic asymmetries, 
namely $\sin^2 \hat{\theta}(m_Z)_{(l)} = 0.23103$.
For the contribution of the first five flavors of quarks to the
$i$-summation in Eq.~(\ref{eq7}) one must invoke dispersion relations and
experimental data on $e^+ \, e^- \rightarrow$~hadrons: we use a
recent update by Marciano \cite{m13}. Further employing $m_W =
80.426\,\mbox{GeV}$, $m_Z = 91.1875\,\mbox{GeV}$, $m_t =
174.3\,\mbox{GeV}$ \cite{m12}, $m_H = 200\,\mbox{GeV}$, we obtain
\bea \kee (0, m_Z) &=& 1.0270 \pm 0.0025 \, , \label{eq8} \\
\rho^{(e ;\, e)} &=& 1.0034\, \label{eq9}.\eea
In the region $115\,\mbox{GeV}\le m_H \le 200\,\mbox{GeV}$, $\rho^{(e;\, e)}$
 varies slowly with $m_H$.
For instance, $\rho^{(e;\, e)} = 1.0037$ at 
$m_H = 115\,\mbox{GeV}$.

Most of the difference between
Eqs.~(\ref{eq6}-\ref{eq9}) and the results reported in
Ref.~\cite{m7} is due to the fact that we have retained the
contributions of the $W-W$ boxes in $\kee (0,m_Z)$ and $\rho^{(e
;\, e)}$ in order to ensure their gauge independence. In contrast,
in Ref.~\cite{m7} these contributions have been separated out from
the form factors in a particular gauge, namely the
't~Hooft-Feynman gauge. In this gauge, the $W-W$ boxes contribute
$\alpha/4 \pi \sms$ to $\rho^{(e ;\, e)}$ and $-\alpha/4 \pi \sms$
to $\kee$. However, 
 in the general $R_\xi$ gauges, they may be arbitrarily
different. A second, smaller difference, is that, as explained
before, we have included the top quark contribution in
Eq.~(\ref{eq7}). We have already pointed out that the $Z-Z$ boxes
are gauge independent and therefore they may be separated out
without affecting the gauge properties of $\kee$ and $\rho^{(e ;\,
e)}$. They are suppressed by a factor $1 - 4 \sms$ and our
calculation of these terms agrees with that reported in
Ref.~\cite{m7}. The contribution of the $Z - \gamma$ boxes, the
diagram of Fig.~\ref{fig2}, and QED corrections are also
proportional to $1 -4 \sms$ and are contained in a function $F_1
(y, Q^2)$ ($y =Q^2/s$; $Q^2 = -q^2$) evaluated in Ref.~\cite{m7}. 
As an interesting illustration, it is pointed out in that paper that
$F_1 (1/2, 0.02\,\mbox{GeV}^2) = -0.0041 \pm 0.0010$.

Putting together the values of the form factors evaluated in the
present paper (Eqs.(\ref{eq8},\ref{eq9})), the $Z - Z$ box
diagrams and the calculation of $F_1 (y, Q^2)$
reported in Ref.~\cite{m7}, one finds that the overall effect of the
electroweak corrections is to replace
\be 1 - 4\,\sin^2 \hat{\theta}_W (m_Z) \rightarrow \rho^{(e
;\,e)}\,\left\{1 - 4\,\kee\,(0, m_Z)\,sin^2 \hat{\theta}_W (m_Z) +
(Z - Z)_{\mbox{box}} + F_1 (y, Q^2) \right\} \, , \label{eq10}\ee
which at $Q^2 = 0.025\,\mbox{GeV}^2$ and $y = 1/2$ equals $0.0454
\pm 0.0023 \pm 0.0010$. This is numerically very close to the
result obtained by Czarnecki and Marciano because these authors chose 
to separate the contributions of the $W-W$ boxes in the 
't~Hooft-Feynman gauge, where they are reasonable
small (cf. Section 4). 
As emphasized in Ref.~\cite{m7}, since $1 -4\,\sin^2 \hat{\theta}_W (m_Z) = 0.07448 \pm
0.00068$, the effect of the electroweak corrections in this case
is to reduce the asymmetries by $\approx 39 \, \%$! Clearly, the
bulk of the reduction is contained in 
\bmath 1 - 4\,\kee\,(0,m_Z)\,\sin^2 \hat{\theta}_W (m_Z) =
0.0495\, . \emath
Detailed studies of radiative corrections to polarized M\o ller scattering
at low and high energies are given in Ref.~\cite{m14} and Ref.~\cite{m15},
respectively. 

It should be pointed out that the correction $\kee (0, m_Z) -1 =
0.0270$ is very different from the corresponding effects in other
processes. For instance, in $\nu_\mu - e$ and $\nu_e - e$
scattering one obtains $\hat{\kappa}^{(\nu_\mu ;\, e)} (0, m_Z) - 1 = -
0.0032$ and $\hat{\kappa}^{(\nu_e ;\, e)} (0, m_Z) - 1 = - 0.0210$,
respectively \cite{m3}. The large difference is mainly due to
sizable ``charge radius'' diagrams that contribute negatively
and to significant and negative $W -W$ and $Z -Z$ box
contributions. In contrast, the vertex diagram of
Fig.~\ref{fig2} as well as the $Z -Z$ boxes are suppressed by $1
-4 \sms$ in the M\o ller scattering case, while the large
corrections associated with the $\gamma - Z$ self-energy are not.

Using the gauge independent form factor $\kee (q^2, \mu)$, it is
possible to define an effective electroweak parameter for $|q^2|
\ll m_W^2$:
\be \see \equiv \mbox{Re}\, \kee (q^2, \mu)\, \sin^2
\hat{\theta}_W (\mu)\, .\label{eq11} \ee
For $q^2 < 0$, $\kee (q^2, \mu)$ is real and the Re instruction is not necessary. 
Dividing Eq.~(\ref{eq11}) by
\be \seff = \mbox{Re}\, \hat{k}(m_Z^2, \mu)\,\sin^2 \hat{\theta}_W
(\mu)\, ,\label{eq12} \ee
where $\hat{k}(m_Z^2, \mu)$ is the form factor discussed in
Ref.~\cite{m5}, and neglecting two-loop effects not enhanced by
powers of $m_t^2$, we find
\be \see = \left\{1+ \mbox{Re}\, \left[\kee (q^2, \mu) -
\hat{k} (m_Z^2, \mu)\right]\ \right\}\, \seff \, . \label{eq14}\ee
The $\mu$ dependence cancels in Eq.~(\ref{eq14}) so that we may
choose $\mu = m_Z$, and Eq.~(\ref{eq14}) becomes
\be \see = \left[\mbox{Re}\, \kee (q^2, m_Z) - 0.00044 \right]\,\seff \, . \label{eq15}\ee
Eq.~(\ref{eq15}) establishes a scale-independent relationship
between the two gauge independent parameters $\see$ and $\seff$, 
which may be regarded as physical observables.

It is worthwhile to point out that the form factors $\hat{\kappa} (q^2, \mu)$ 
and $\hat{k} (m_Z^2, \mu)$ are quite different conceptually
even when  $\hat{\kappa}$ is evaluated at $q^2 = m_Z^2$. 
While $\hat{\kappa}$ is relevant to
four-fermion scattering processes, $\hat{k} (m_Z^2, m_Z)$
involves the decay amplitude of an on-shell $Z^0$ into an $l -
\overline{l}$ pair.

\boldmath
\section{The running of $\sin^2 \hat{\theta}_W$}
\unboldmath

In this Section we discuss the possibility of constructing a
running electroweak mixing parameter for arbitrary values of
$q^2$. We will impose four theoretical requirements:
i) it should be process independent ii) since $\sin^2 \hat{\theta}_W$ is related to $\gamma
-Z$ mixing, it should involve the $A_{\gamma Z} (q^2)$ self-energy
in a fundamental way
iii) it should be gauge independent, at least in the class
of $R_\xi$ gauges
iv) it should be as simple as possible.

At first hand, these requirements seem difficult to satisfy. In
fact, we have seen in Section~2 that, in order to obtain gauge
independent form factors, the contributions of box diagrams must be
included. Since in general box diagrams depend on two kinematic
variables, $s$ and $q^2$, and are process dependent, it is not
trivial to see how to achieve our aims. However, there is a well
known framework that allows us to satisfy these conditions, namely
the Pinch Technique (PT) \cite{m16,m17,m18,m19,m20,n2,n3,m21,m22,m23}. 
We recall that the PT is a
prescription that judiciously combines the conventional
self-energies with ``pinch parts'' from vertex and box diagrams
in such a manner that the new self-energies are gauge independent
and possess desirable theoretical properties. In
Ref.~\cite{m20}, it was shown that the ``pinch parts'' can be identified with
amplitudes involving appropriate equal-time commutators of currents,
which explains why they are process independent and unaffected by
strong interaction dynamics.

In this Section we discuss a running electroweak mixing
parameter $\sin^2 \hat{\theta}_W (q^2)$ defined in terms of the PT
$\gamma Z$ self-energy. Specifically,
\be \sin^2 \hat{\theta}_W (q^2) \equiv \left(1 -
\frac{\hat{c}}{\hat{s}}\, \frac{a_{\gamma Z}(q^2,
\mu)}{q^2}\right)\,\sin^2 \hat{\theta}_W (\mu) \,, \label{eq16}\ee
where $a_{\gamma Z}(q^2, \mu)$, the PT $\gamma Z$ self-energy of
the SM, can be conveniently expressed as \cite{m20}
\be a_{\gamma Z}(q^2, \mu) = A_{\gamma Z}(q^2, \mu)|_{\xi_W = 1} -
\frac{2 e^2}{\hat{c} \hat{s}}\,\left(2\,q^2\,\cms -m_W^2
\right)\,I_{W W}(q^2, \mu) \,. \label{eq17} \ee
In Eq.~(\ref{eq17}) $A_{\gamma Z}(q^2, \mu)|_{\xi_W = 1}$ is the
conventional $\gamma-Z$ self-energy evaluated in the
't~Hooft-Feynman gauge $\xi_W = 1$, and
\be I_{W W}(q^2, \mu) = \frac{1}{16 \pi^2}\, \int_0^1 dx\,
\ln{\left[\frac{m_W^2 - q^2\,x\,(1-x) -i\,\epsilon}{\mu^2}
\right]} \,. \label{eq18}\ee
Very simple analytic formulae for $I_{W W}(q^2, \mu)$ are given in
Eqs.~(A5 - 7) of Ref.~\cite{m20}. In Eq.~(\ref{eq18}) we have
performed  the $\overline{\mbox{MS}}$ subtraction of 
$\delta = (n -4)^{-1} + (\gamma -\ln{4 \pi})/2$. It is
understood that the same subtraction has been implemented in
$A_{\gamma Z}(q^2, \mu)|_{\xi_W = 1}$.

Since the r.h.s. of Eq.~(\ref{eq16}) is process and gauge
independent, it satisfies our theoretical requirements. It is also
important to remember that, unlike $A_{\gamma Z}(q^2, \mu)|_{\xi_W =
1}$, $a_{\gamma Z}(0, \mu) = 0$, so that Eq.~(\ref{eq16}) is
regular as $q^2 \to 0$.

It is convenient to define
\be \hat{\kappa}^{PT}(q^2, \mu) = 1 -
\frac{\hat{c}}{\hat{s}}\,\frac{a_{\gamma Z} (q^2, \mu)}{q^2} \,.
\label{eq19} \ee
In the range $|q^2| \ll m_W^2$, we find
\bea \hat{\kappa}^{PT}(q^2, \mu) &=& 1 + \frac{\alpha}{2 \pi
\hat{s}^2}\,\ln{\left(\frac{m_Z}{\mu}\right)}\left[-\frac{1}{3}\,\sum_i\,(C_i\,Q_i
-4\,\sms\,Q_i^2) + 7\,\cms +\frac{1}{6}\right] \nonumber \\ &+&
\frac{\alpha}{2 \pi \hat{s}^2}\,\left[-\sum_i (C_i\,Q_i
-4\,\sms\,Q_i^2)\,I_i(q^2)+ \left(\frac{7}{2}\,\cms +
\frac{1}{12}\right)\,\ln{c^2}  -\frac{\cms}{3} \right]\,
,\label{eq20}
 \eea
Comparing Eq.~(\ref{eq20}) with Eq.~(\ref{eq5}), we have, for
$|q^2| \ll m_W^2$:
\be \hat{\kappa}^{PT} (q^2, \mu) = \kee (q^2, \mu)
+\frac{17}{18}\,\frac{\alpha}{2 \pi \sms}\, . \label{eq21} \ee
Thus, while $\kee (0, m_Z) = 1.0270 \pm 0.0025$ (cf.
Eq.~(\ref{eq8})),
\be \hat{\kappa}^{PT} (0, m_Z) =  1.0317 \pm 0.0025\,, \label{eq22} \ee
a difference of $4.7 \times 10^{-3}$. We see that the PT form
factor approximates rather well $\kee(0, m_Z)$ at $q^2=0$. 
More interestingly, the PT running parameter evaluated at $q^2 =0$ almost exactly
absorbs the complete calculation reported in Eq.~(\ref{eq10}) for
$y =1/2$ and $Q^2 = 0.025\,\mbox{GeV}^2$. In fact, using
Eqs.~(\ref{eq16},\ref{eq19},\ref{eq22}), we have
\be \sin^2 \hat{\theta}_W (0) = 0.2387 \pm 0.0006 \, .
\label{eq23} \ee
This leads to
\be 1 -4\,\sin^2 \hat{\theta}_W (0) = 0.0452 \pm 0.0023 \,,
\label{eq24} \ee
in very close agreement with $0.0454 \pm 0.0023 \pm 0.0010$,
reported after Eq.~(\ref{eq10}),  when all electroweak corrections
are taken into account. Of course, this very accurate agreement
will generally not hold for other values of $y$ and $Q^2$, but
it is interesting that $1 -4\,\sin^2 \hat{\theta}_W (0)$ does
absorb quite precisely the bulk of the corrections to M\o ller
scattering at $s \ll m_W^2$.

In order to evaluate $\seq$ as a function of $q^2$, we set $\mu = m_Z$ and 
employ the expressions for $A^{(f)}_{\gamma Z} (q^2, m_Z)$ from
Ref.~\cite{m1}, $A^{(b)}_{\gamma Z} (q^2, m_Z)|_{\xi_W = 1}$ from
Ref.~\cite{m24} ($f$ and $b$ mean fermionic and bosonic
contributions), and $I_{W W} (q^2, m_Z)$ from Ref.~\cite{m20}.
As for $\alpha$, we follow the approach of Ref.~\cite{m25} and replace
$\alpha^{-1} \to \alpha^{-1}(q^2)$,
\be
\alpha^{-1}(q^2) = \alpha^{-1} -\frac{2}{\pi}\sum_i Q_i^2 \int_0^1 dx\, x\,(1-x)\,
\ln{\left\{\frac{m_i^2 - q^2\,x\,(1-x)}{m_i^2}\right\}}\, . 
\ee
In evaluating  $A^{(f)}_{\gamma Z} (q^2, m_Z)$ and $\alpha^{-1}(q^2)$, we employ
the effective light quark masses and the QCD correction factor discussed in Ref.~\cite{m13}.

\begin{figure}[!ht] 
\centering
\psfrag{m}{ {\small $Q\ \left[ \mbox{GeV} \right]$}}
\psfrag{s}{{\footnotesize $\seq$ }}
\resizebox{12cm}{7cm}{\includegraphics{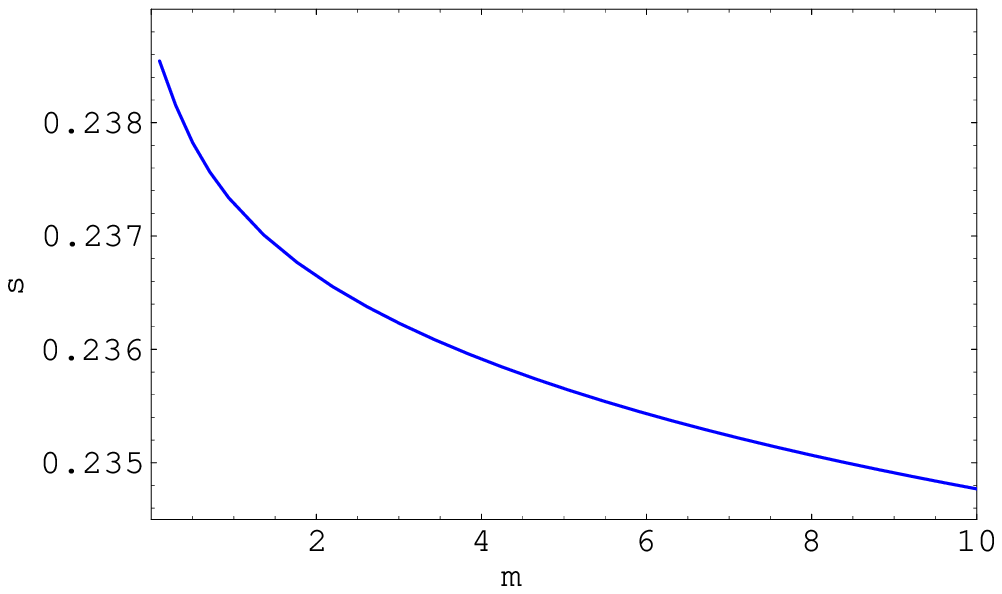}}
\caption{$\seq$ as a function of $Q = (-q^2)^{1/2}$ in the space-like domain  
$q^2 < 0$, for $0 \le Q \le 10\ \mbox{GeV}$.}
\label{fig3}
\end{figure}

\begin{figure}[!ht] 
\centering
\psfrag{m}{ {\small $Q\ \left[ \mbox{GeV} \right]$}}
\psfrag{s}{{\footnotesize\ $\seq$ }}
\resizebox{12cm}{7cm}{\includegraphics{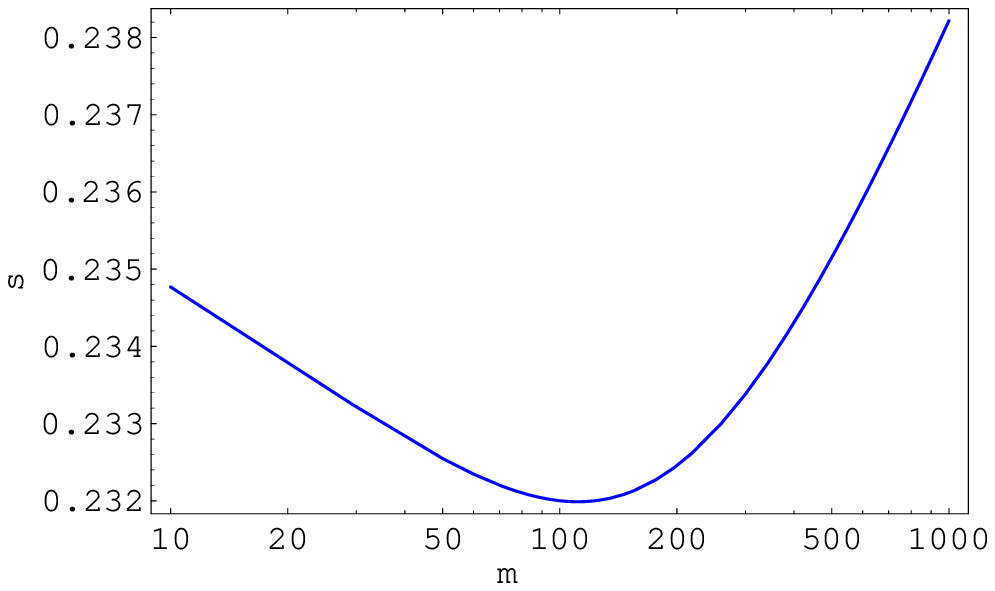}}
\caption{Same as in Fig.~\ref{fig3}, for $10\ \mbox{GeV} \le Q \le 1\ \mbox{TeV}$. }
\label{fig3b}
\end{figure}

Figs.~\ref{fig3} and \ref{fig3b} show $\seq$ in the space-like domain $q^2 < 0$, 
appropriate for  $e^-$-$e^-$ colliders, as a function of $Q = (-q^2)^{1/2}$.
Table~\ref{tab1}
gives a few representative values of $\hat{\kappa}(q^2, m_Z)$ and
$\sin^2 \hat{\theta}_W (q^2)$ in that region.

We see that $\sin^2
\hat{\theta}_W (q^2)$ equals $0.2387$ at $Q =0$, $0.2320$ at $Q =
m_Z$, reaches a minimum of $0.23199$ at $Q = 111 \, \mbox{GeV}$,
and then increases monotonically to $0.2352$ at $Q = 500 \,
\mbox{GeV}$ and $0.2382$ at $Q = 1\, \mbox{TeV}$. 

Fig.~\ref{fig3b} bears a close resemblance to a curve presented
in Refs.~\cite{m8,m9} for a running parameter constructed on
the basis of the diagrams in Figs.(1a,d). The
theoretical foundation of the two running parameters is, however,
very different. While the PT self-energy is gauge independent within the 
class of $R_\xi$ gauges, the
sum of the diagrams in Figs.(1a,d) is not. Thus, in
principle, by varying the gauge in the second approach one can
alter the values and $Q^2$ dependence of the running parameter.

A second problem, already discussed in Ref.~\cite{m20}, is that diagram (1d)
is not truly process-independent, even in the limit of neglecting the external fermion
masses. For instance, when the external fermion is a quark or a hadron, there are QCD
corrections not present in the leptonic case. As explained in Ref.~\cite{m20}, this 
problem is neatly bypassed in the PT approach since the pinch part of Fig.(1d) is
unaffected by strong interaction dynamics.

In the time-like domain $q^2 > 0$, $\kpt$ is complex and 
we define the running parameter as 
\be
\seq \equiv \mbox{Re}\left[\kptm  \right] \sin^2 \hat{\theta}_W
(m_Z)
\ee

\begin{figure}[!ht] 
\centering
\psfrag{m}{ {\small $Q\ \left[ \mbox{GeV} \right]$}}
\psfrag{s}{{\footnotesize $\seq$ }}
\resizebox{12cm}{7cm}{\includegraphics{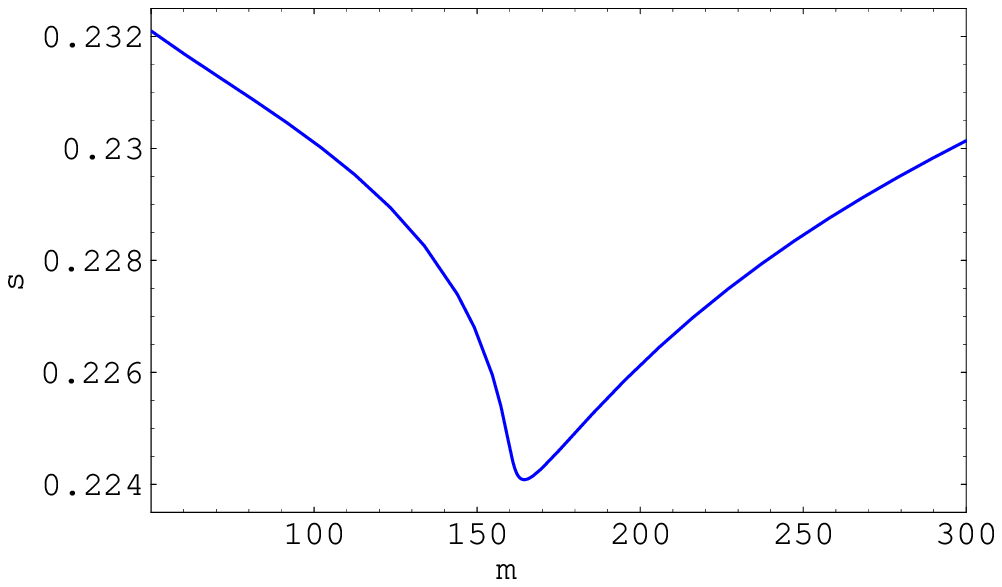}}
\caption{$\seq$ as a function of $Q = (q^2)^{1/2}$ in the time-like domain  
$q^2 > 0$, for $50 \le Q \le 300\ \mbox{GeV}$.}
\label{fig4}
\end{figure}

\begin{figure}[!ht] 
\centering
\psfrag{m}{ {\small $Q\ \left[ \mbox{GeV} \right]$}}
\psfrag{s}{{\footnotesize\ $\seq$ }}
\resizebox{12cm}{7cm}{\includegraphics{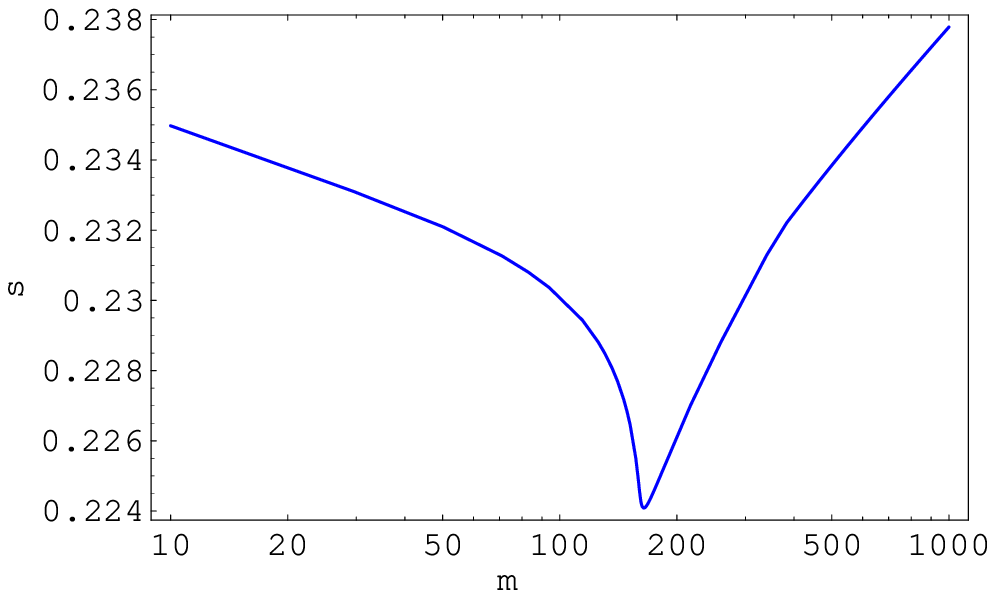}}
\caption{Same as in Fig.~\ref{fig4}, for $10\ \mbox{GeV} \le Q \le 1\ \mbox{TeV}$.}
\label{fig4b}
\end{figure}

In Figs.~\ref{fig4} and \ref{fig4b} we present values of $\sin^2 \hat{\theta}_W
(q^2)$  in the time-like domain $q^2 > 0$, appropriate to $e^+
-e^-$ colliders, as a function of $Q = (q^2)^{1/2}$.
 A sharp decrease associated with the $W -W$
threshold is very visible. In order to soften the behavior in that
region we have included the $W$ width by means of the replacement
$m_W^2 \rightarrow m_{2\, W}^2 - i\,m_{2\, W}\,\Gamma_{2\, W}$, with
$m_{2\, W} = m_W / (1  + (\Gamma_W / m_W)^2)^{1/2} = 80.398\,
\mbox{GeV}$ and $\Gamma_{2\, W} = m_{2\, W} \Gamma_W/ m_W = 2.117 \,
\mbox{Gev}$ \cite{m26}.

At $Q = m_Z$, we find $\sin^2 \hat{\theta}_W (m_Z^2) = 0.23048$,
which is lower than $\seff = 0.23148$ by $0.43 \, \%$. Although
not in perfect consonance, the two parameters are rather close.
Other representative values of $\hat{\kappa}^{PT}$ and $\sin^2
\hat{\theta}_W (q^2)$ in the time-like region are given in
Table~\ref{tab2}. We see that $\sin^2 \hat{\theta}_W (q^2)$ ranges
from $0.2387$ at $Q =0$ to $0.2305$ at $Q = m_Z$, reaches a
minimum of $0.2241$ at $Q = 164\,\mbox{GeV}$, and then increases
monotonically to $0.2338$ at $Q = 500 \, \mbox{GeV}$ and $0.2378$
at $Q = 1 \, \mbox{TeV}$.

It is interesting to note that $\sin^2 \hat{\theta}_W (- m_Z^2)$ is slightly
larger then $\seff$, the difference being  $0.22 \, \%$.
Thus, $\seff$ lies between  $\sin^2 \hat{\theta}_W (m_Z^2)$ and  
$\sin^2 \hat{\theta}_W (- m_Z^2)$.

\section{Discussion}

Aside from the general observation that physical results in gauge theories
should be parametrized in a gauge-independent manner, there are specific reasons
why this is particularly important in the case of the electroweak form factors
$\rho$ and $\hat{\kappa}$:

i) If $\kee (q^2, \mu)$ is gauge independent, 
the effective electroweak mixing parameter $\see$ defined in Eq.(\ref{eq11})
is also gauge independent, and consequently may be regarded as a physical
observable. In particular, $\seezero$ can be measured by
polarized M\o ller scattering with considerable precision. 
If $\kee (q^2, \mu)$
is defined in a gauge-dependent manner, this is theoretically
unfounded, since  $\see$ would not qualify as a
physical observable.

ii) The parameterization in terms of $\rho$ and $\hat{\kappa}$ involves a
factorization of one-loop electroweak corrections (see, for example,
Eqs.~(\ref{eq3},\ref{eq3bis},\ref{eq10})). 
If the form factors are defined in a gauge-dependent manner,
one can make the two-loop effects induced by the factorization to vary
arbitrarily by simply changing the gauge used to calculate the form
factors. 
These effects can be very large for large values of the gauge
parameter and, in fact, the calculation diverges in the unitary, i.e. the
physical, gauge!

iii) It should be also pointed out that the electroweak form factors for other
processes, such as discussed in Refs.\cite{m1,m2,m3}, have been defined in a
gauge-independent manner.

In order to circumvent these theoretical problems,
in Section~2 we have discussed the derivation of gauge
independent form factors $\rho^{(e ;\, e)}$ and $\kee (q^2, \mu)$
appropriate to M\o ller scattering at $s \ll m_Z^2$.
For the reasons explained after Eq.~(\ref{eq10}), the overall electroweak
corrections including these form factors, as well as the gauge-independent
corrections that have been separated out, agree numerically very closely
with the results of Ref.~\cite{m7}.
We have pointed out that $\kee (q^2, \mu) -1$ is quite different
from the corresponding form factors in other processes such as
$\nu_\mu - e$ or $\nu_e - e$ scattering. Thus, it is not possible,
even at $|q^2| \ll m_W^2$, to find a universal electroweak mixing
parameter that absorbs the bulk of the electroweak corrections in
all processes.

However, as emphasized in Refs.~\cite{m7,m8,m9}, experiments on polarized M\o
ller scattering are very special in that the asymmetries measured
in that process are greatly affected by $\kee (q^2, m_Z)$
and may be used to measure the effective mixing parameter $\see$. At
$s \ll m_W^2$ this is of considerable present interest in view of the proposed E158 
experiment at SLAC \cite{e158}.

In Section~2 we have also derived a scale-independent relation between $\see$
and $\seff$. In particular, this relation may be employed to discuss 
the electroweak corrections
to M\o ller scattering in the effective scheme of renormalization, in which
residual scale dependencies cancel in finite orders of perturbation theory, and $\seff$ 
plays the role of the basic electroweak parameter \cite{m27,m28,m29,m30}.

In Section~3 we have discussed a gauge and process independent
running parameter based on the PT $\gamma Z$ self-energy \cite{m20}. 
At $q^2=0$ it absorbs very precisely the electroweak corrections 
to M\o ller scattering evaluated in Ref.~\cite{m7} at $y = 1/2$
and $Q^2 = 0.025\, \mbox{GeV}^2$, and at $q^2 = m_Z^2$ it lies
within $0.43 \, \%$ of $\seff$. Thus it provides an attractive
theoretical and phenomenological framework to describe the running of
the electroweak mixing parameter over a large range of $q^2$ values.


\section*{Acknowledgments}

The work of A.~S. was supported in part by NSF Grant PHY-0245068.
The work of A.~F. was supported by the DFG-Forschergruppe 
\emph{``Quantenfeldtheorie, Computeralgebra und Monte-Carlo-Simulation''}.

\section*{Appendix}

In this Appendix we report the numerical results obtained if one employs as input 
the central value of $\seff$ derived from the leptonic asymmetries, namely $\seff = 0.23113$, rather 
than the world average. 

Without decoupling the top quark contributions, we have $\sin^2 \hat{\theta}_W (m_Z) = 0.23103$. 
The value in Eq.~(\ref{eq8}) is replaced by $1.0271 \pm 0.0025$, while  Eq.~(\ref{eq10}) equals
$0.0467 \pm 0.0023 \pm 0.0010$. \mbox{$1 - 4\,\hat{\kappa}^{(e; \,e)} (0, m_Z)\,\sin^2 \hat{\theta}_W (m_Z)$}
 becomes $0.0508$,
to be compared with  $1 - 4\,\sin^2 \hat{\theta}_W (m_Z) = 0.0759$. 
Eqs.~(\ref{eq22},\ref{eq23},\ref{eq24}) equal
$1.0319 \pm 0.0025$, $0.2384 \pm 0.0006$, and $0.0464 \pm 0.0023$, respectively. 
The values of $\sin^2 \hat{\theta}_W (q^2)$ are smaller by $0.0003$ or $0.0004$ that those
in Table~\ref{tab1} and by $0.0004$ than those in Table~\ref{tab2}.

\newpage
\vspace*{3cm}


\begin{table}[!ht] 
\begin{center}
\begin{tabular}{|c||c|c|}
\hline
& 	 &  \\
$Q \left[ \mbox{GeV} \right] $ & $\kptm$ & $\seq$ \\
&  	&   \\
\hline \hline
&  	&  \\ 
0 & 1.0317  &  0.2387  \\
&  	&  \\
\hline
&  	&  \\
$m_Z$ & 1.0028 & 0.2320  \\
&  &  \\
\hline
&  &  \\
111& 1.0026 & 0.2320 \\
&  &  \\
\hline
&  &  \\
500& 1.0163 & 0.2352 \\
&  &  \\
\hline
&  &  \\
1000& 1.0296 & 0.2382 \\
&  &  \\
\hline
\end{tabular} 
\caption{$\kptm$ and $\seq$ for $q^2 < 0$ at different values of $Q$ ($Q = \sqrt{-q^2}$).} 
\label{tab1}
\end{center}
\end{table}

\begin{table}[!ht]  
\begin{center}
\begin{tabular}{|c||c|c|}
\hline
& 	 &  \\
$Q \left[ \mbox{GeV} \right] $ & $\kptm$ & $\seq$ \\
&  	&   \\
\hline \hline
&  	&  \\
$m_Z$ &  0.9961 & 0.2305 \\
&  &  \\
\hline
&  &  \\
164&  0.9685 & 0.2241 \\
&  &  \\
\hline
&  &  \\
500 & 1.0107 & 0.2338 \\
&  &  \\
\hline
&  &  \\
1000 & 1.0277 & 0.2378 \\
&  &  \\
\hline
\end{tabular} \caption{$\kptm$ and $\seq$ for $q^2 > 0$ at different 
values of $Q$ ($Q = \sqrt{q^2}$).} \label{tab2}
\end{center}
\end{table}

\end{document}